\documentclass[12pt,preprint]{aastex}
\usepackage{emulateapj5}

\newcommand{\gll}{G11.2$-$0.3 }

\begin{document}

\title{The Expansion of G11.2$-$0.3, a Radio Composite Supernova
  Remnant}

\author{Cindy Tam and Mallory S. E. Roberts \altaffilmark{1}}

\affil{Department of Physics, Ernest Rutherford Physics Building,
McGill University, 3600 University Street, Montreal, Quebec,
H3A 2T8, Canada; tamc@physics.mcgill.ca, roberts@physics.mcgill.ca}

\altaffiltext{1}{Department of Physics and Center for Space Research,
Massachusetts Institute of Technology, Cambridge, MA 02139}

\begin{abstract}
We compare recent observations of the supernova remnant \gll taken
with the VLA during 2001$-$02 with images from VLA archives
(1984$-$85) to detect and measure the amount of expansion that has
occurred during 17 years.  The bright, circular outer shell shows a
mean expansion of ($0.71 \pm 0.15$)\% and ($0.50 \pm 0.17$)\%,
from 20- and 6-cm data, respectively, which corresponds to a rate of
$0\farcs057 \pm 0\farcs012$/yr at 20 cm and $0\farcs040 \pm
0\farcs013$/yr at 6 cm. From this result, we estimate the age of the
remnant to be roughly between 960 and 3400 years old, according to
theoretical models of supernova evolution.  This is highly inconsistent
with the 24000 yr characteristic age of PSR~J1811$-$1925, located at
the remnant's center, but, rather, is consistent with the time since
the historical supernova observed in 386~AD.  We also predict that
\gll is currently in a pre-Sedov evolutionary state, and set
constraints on the distance to the remnant based on Chandra X-ray
spectral results.  
\end{abstract}

\keywords{pulsars: individual (PSR J1811$-$1925) --- supernovae:
  individual (G11.2$-$0.3) --- supernova remnants}

\section{INTRODUCTION}

The process of supernova remnant expansion evolution has long been the
subject of thorough investigation, and as a result we can be confident
of a few well established facts.  In the initial free expansion stage,
a tremendous amount of energy in the form of material ejecta is thrown
outwards in a (assumed) spherically symmetric explosion, driving a
shock wave into the ambient medium.  Later, when the mass of the
swept-up material considerably exceeds the ejected material mass, the
supernova remnant enters the Sedov stage \citep{sedo93}, as a reverse
shock reaches the center of the remnant and the forward shock
undergoes significant deceleration.  This simple picture is
complicated by the presence of a pulsar wind nebula expanding within,
but separate from, the SNR bubble \citep{chev84,reyn84}.  It
eventually encounters the reverse shock, which induces complicated
radial reverberations, before relaxing into Sedov phase expansion
\citep{vand01,blon01}. Each stage of evolution can be described by the
expansion parameter $m$, defined by $R \propto t^m$, where $R$ is the
linear radius and $t$ is the remnant age.
\par
In G11.2$-$0.3, we find a textbook example of a composite supernova
remnant (SNR) comprising an extremely circular radio and X-ray shell,
a pulsar wind nebula (PWN) contained within the SNR
\citep{vasi96,koth01}, and an X-ray pulsar PSR~J1811$-$1925 
\citep{tori97} located at the center. Its high surface brightness and
the extremely centralized position of the pulsar within the remnant
imply that it is much younger than the pulsar's characteristic age,
$\tau = 24 000$ yr \citep{tori99,kasp01}, and support the possible
association with the historical supernova (SN) event of 386~AD
\citep{clar77}.  Furthermore,  this discrepancy suggests the pulsar's
current spin period is very near its initial value and that its
spin-down energy, $\dot E = 6.4 \times 10^{36}$ ergs/s, has remained
nearly constant since the supernova explosion.
\par 
\gll is an ideal SNR for detailed study due to the observability of
emission from all of its components, either at hard X-ray, thermal
X-ray or radio frequencies; for a description of its radio and X-ray
properties see \citet{tamr02} and \citet{robe03}.  The purpose of
performing an expansion measurement is to set unambiguous upper and
lower limits on its age by examining the expected behaviour of the
outer shock during the free expansion (high velocity) and Sedov
(low velocity) phases.  

\section{OBSERVATIONS}

We obtained 20- and 6-cm data of \gll taken during 1984$-$85 (epoch 1)
from the VLA archival database.  Details of these observations can be
found in Table 1 of \citet{tamr02}; data from 1985 February were
omitted from this analysis due to poor calibration.  Our recent
observations made during 2001$-$02 (epoch 2) at 20 and 6 cm (1465 and
4860 MHz, respectively) are outlined in Table \ref{tab:obs}.
\par
The data processing was performed using standard
procedures within the MIRIAD package \citep{saul99}, in mosaic and
multi-frequency synthesis mode.  We performed calibration and editing
on each data set individually, before combining all the data of a
particular frequency band and epoch.  The primary gains were
determined using 3C~286 and 3C~48, and phase calibrations were made
from observations of 1743$-$038, 1751$-$096, 1751$-$253 and 1820$-$254
(J2000).  Imaging was performed with Robust weighting \citep{saul99}
as a compromise between maximized signal-to-noise and resolution.  We
utilized the maximum entropy method (MEM) algorithm for deconvolution
\citep{corn99}, and applied self-calibration iteratively to improve
phase and amplitude calibrations.

\section{ANALYSIS}

The VLA and other radio interferometers have been used to measure the
expansion of many SNR such as G11.2$-$0.3.  For a summary of remnants
and the techniques used to study them, see \citet{reyn97} and the
references therein.  Because it is not always possible to make new
observations with the same $u-v$ coverage as the archival data, it 
is important to match up as many of the properties which might affect
the quality of our results as best possible, before attempting to
directly compare the final images from each epoch.  We used the final
self-calibrated epoch 2 clean map, after correcting 
it for primary beam attenuation, as the model for self-calibrating
epoch 1 data, as described by \citet{mass86} in his cross-calibration
method.  The purpose of this step was to apply the same residual
calibration errors to epoch 1 as existed in epoch 2, thereby minimizing
the effects of self-calibration errors on the final subtracted map.
In order to match the spatial scales of our images at both epochs, we
used MIRIAD modelling procedures to spacially filter the $u-v$ coverage
of epoch 2 data to match that at epoch 1, thus creating an epoch 2
dataset with degraded visibility coverage \citep{gaen99}. 
\par
Rather than subtract our maps in the $u-v$ plane as done by
\citet{mass86}, we instead mimicked the direct approaches of
\citet{moff93} and \citet{reyn97} (used to measure the remnant
expansions of SN~1006 AD and Tycho's SN (3C~10), respectively) who
adopted procedures outlined by \citet{stro82}, \citet{tang85} and
\citet{dick88}.  Using the MIRIAD task IMDIFF, we fit the five
parameters, described by the maximum likelihood algorithm of
\citet{tang85}, between our final images: expansion, amplitude
(the mean brightness ratio between epochs), x-shift (in the negative
right ascension direction), y-shift (declination), and DC offset (the
difference in background brightness levels).  The best-fit geometrical
center of the shell was found to be $< 0\farcs5$ from the position of
PSR J1811$-$1925.  We then fixed the amplitude, x-shift, y-shift and
offset at the fitted values, performed a series of image
subtractions of epoch 1 from epoch 2, each time artificially scaling
the epoch 1 image by an expansion factor between 0\% and 1.5\% in
steps of 0.1\%, and examined the radial profiles of each difference
image for the best-fit expansion factor, or more specifically, the
profile that most resembled a line of zero slope.  It was evident when
examining the difference images that 1.5\% was a sufficiently large
upper bound on the expansion factor.  The difference images were
convolved with Gaussians whose FWHM were the same as the synthesized
beams' ($19\farcs3 \times 15\farcs3$ at 20 cm, $8\farcs3 \times
7\farcs7$ at 6 cm). Figure~\ref{fig:dif} shows a map of epoch 2 minus
epoch 1 at 20 cm with zero expansion.
\par
We divided the remnant into 24 wedge-shaped regions of $15\degr$
azimuthally, and found the average flux of each difference image in
annular ring sections as a function of radius between 1$'$ and 3$'$,
recalling that the radius of \gll is $\sim2\farcm25$ \citep{tamr02}.
The purpose of averaging over wedges, as opposed to simply taking
radial cuts, was to smooth out small fluctuations that might have
corrupted our data \citep{stro82}.  The next task was to determine
which of these profiles of average residual flux vs. radius most
resembled a flat line.  A straightforward $\chi^2$ analysis was not
possible, due to our lacking measured uncertainties $\sigma$ on the
residual profile data points.  Therefore, we chose to take a modified
approach to $\chi^2$-fitting in the hopes of obtaining somewhat
legitimate error bars.  We weighted each profile data point according
to the total intensity in that region at that radius, in lieu of
$\sigma$, and calculated $\chi^2$ at each expansion value using a flat
line as the expected difference profile.  To find the best expansion
and uncertainties for a particular region, we divided the $\chi^2$
values by the minimum $\chi^2$ of that region; therefore, when we
fitted the values to a parabola, its minimum was forced to be near 1.
The best expansion was determined by the location of the minima, and
the $1\sigma$ errors from the location of $\chi^2 = 2$. It should be
noted that although these error bars do not represent $1\sigma$ in the
traditional sense, they do provide a rough indication of each
measurement's precision. 

\section{RESULTS}

\begin{figure}
\epsscale{0.5}
\plotone{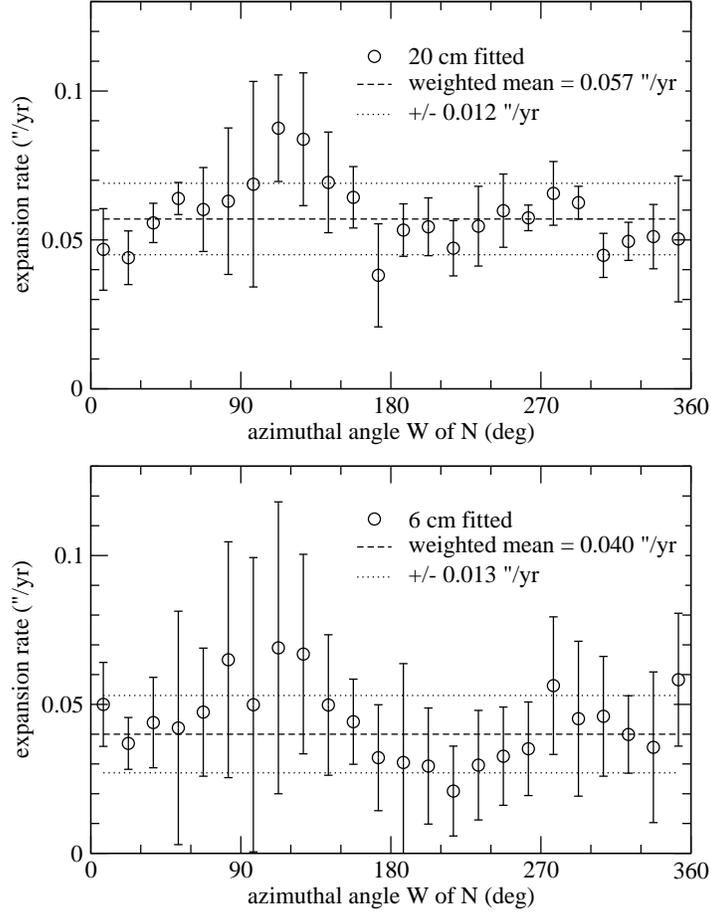}
\figcaption[f2.eps]{Azimuthal variation of average expansion rate,
  from 20- (top) and 6-cm (bottom) data.  The fitted values (circles)
  determined by a pseudo $\chi^2$ analysis with $\sim1\sigma$ error
  bars are shown (solid lines), as is the weighted mean (dashed line)
  and its RMS scatter (dotted lines) for comparison.\label{fig:exp}} 
\end{figure}

Figure~\ref{fig:exp} contains a plot of the best expansion rate
estimates as a function of azimuthal angle.  As expected, the largest
uncertainies correspond to the regions of the shell that are most
diffuse, specifically the south-western quadrant (between $90\degr$
and $180\degr$ W of N).  We calculate the weighted mean expansion rate
for the entire shell, and find the overall expansion during the
$\sim17$ yr period separating 
the epochs to be ($0.71 \pm 0.15$)\% from 20-cm data and ($0.50 \pm
0.17$)\% from 6-cm data.  This corresponds to a rate of $0\farcs057
\pm 0\farcs012$/yr and $0\farcs040 \pm 0\farcs013$/yr (from 20- and
6-cm data, respectively).  The uncertainties represent the RMS
deviation about the weighted mean.  The expansion parameter $m \simeq
(\Delta R/R)/(\Delta t/t)$ will later be used to constrain a rough
estimate of the age of G11.2$-$0.3; here, we assume that the age is
equivalent to the time since SN 386~AD, $t=1616$ yr, and calculate
$m=0.68 \pm 0.14$ (20 cm) and $0.48 \pm 0.16$ (6 cm).  We do not
know why there exists a general trend in the difference between the
two sets of results, making the expansion at 20 cm appear consistently
greater than that at 6 cm; however, we note that the 6-cm data is more
susceptible to errors in primary beam correction and incomplete $u-v$
coverage.  Even so, the error bars on our data points overlap
significantly and the calculated mean expansion values agree within
our scatter-based uncertainties.
\par
As an independent check, we compare our measured expansion with the
general overall expansion for the entire remnant as fit by IMDIFF.
Our results agree with the fit values of 0.7\% and 0.5\% at 20- and
6-cm, respectively.

\section{DISCUSSION}

In the first phase of SN evolution, the outer shock wave freely
expands into the surrounding medium such that $R \propto t$, and the
mass of the ejected material $M_{ej}$ is much greater than the mass of
swept up material $M_{sw}$.  As the forward shock begins to
decelerate, a reverse shock is driven towards the center of the SNR 
where it interacts with the PWN, if present
\citep{reyn84,blon01,vand01}.  Eventually these interactions dissipate  
as the SNR enters the Sedov phase.  At this stage, $M_{ej} << M_{sw}$,
and the expansion is described by the Sedov solution $R \propto
t^{0.4}$ \citep{sedo93}.  Interactions between the SN shock front and
its ambient medium, which has a density profile thought to be affected
by a circumstellar wind produced by pre-supernova mass loss of Type~II
SN, are currently the subject of extensive investigations
\citep{chev82,chev94,true99}. The $m$ values predicted for such an
environment tend to lie between the extreme values of free expansion
and the classical Sedov solution for a constant density medium.
Therefore, we will consider these classical definitions for now. 
\par
Assuming that \gll has been in either the free expansion or Sedov
state continuously since birth, we use our above results and the
definition of $m$ to calculate upper and lower constraints on its age,
where $\Delta R/R = \Delta \theta/\theta = (0.0071 \pm
0.0015)$ and (0.0050 $\pm$ 0.0017).
From our 20-cm result, we find upper and lower age limits,
corresponding to $m=1$ and $m=0.4$ respectively: $t_{\uparrow} = 2400
\pm 500$~yr, $t_{\downarrow} = 960 \pm 200$~yr.  Likewise, from the
6-cm result we find $t_{\uparrow} = 3400 \pm 1100$~yr and
$t_{\downarrow} = 1400 \pm 500$~yr.  It is immediately obvious that
the result of this calculation is highly inconsistent with the
characteristic age of PSR~J1811$-$1925, $\tau = 24 000$ yr; it is,
however, of the same order of magnitude as the time passed since SN
386 AD, and indeed 1616~yr falls easily within these constraints.  To
eliminate any doubt as to the association between the pulsar and SNR,
see the detailed discussion by \citet{kasp01}; also discussed are the
implications of this age discrepancy on young pulsar astronomy.
\par
\begin{figure}
\epsscale{0.5}
\plotone{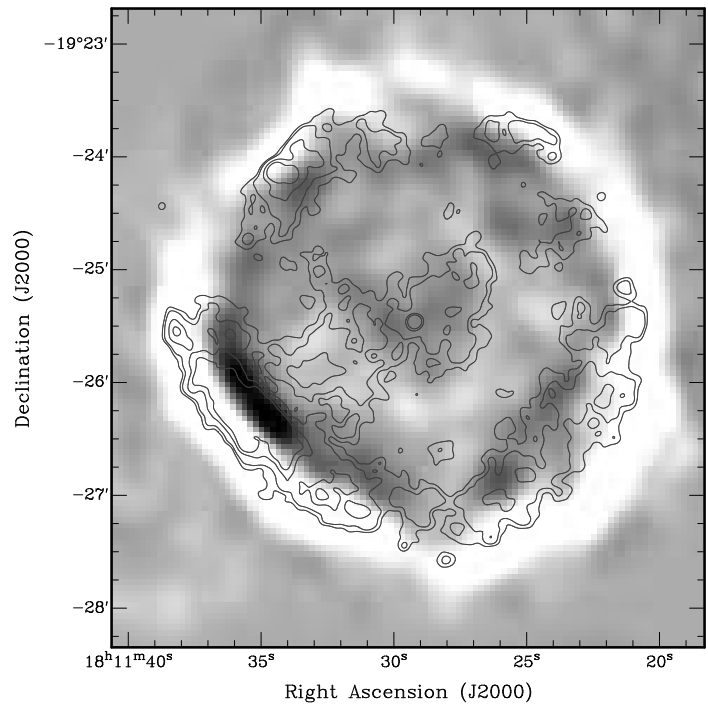}
\figcaption[f1.eps]{20-cm difference map of epoch 1 image subtracted
  from epoch 2 image, with no expansion applied.  Positive (white)
  emission on the outer shell and negative (black) emission on the
  inner shell indicate that a noticeable amount of expansion has
  occured between epochs. Contours of the Chandra 0.6$-$1.65 keV X-ray
  image, smoothed with a 5$''$ Gaussian, are shown at levels of 1.5,
  2, 3.5, and $5\times 10^{-6}$ photons/cm$^2$/s/pixel. The extended
  structure in the interior of the X-ray shell is thought to be the
  PWN forward shock; this is discussed in
  \citet{robe03}.\label{fig:dif}}
\end{figure}

Although the PWN is not bright enough to measure a rate of expansion
in a similar fashion, we find the presence
of flux outlining the general region of the PWN in the zero expansion
difference image noteworthy (see Figure~\ref{fig:dif}).  The positive
emission is indicative of positive expansion; therefore, we conclude
that the reverse shock is not compressing the PWN shock front and the
SNR is not yet in the Sedov phase.  We look to the study of
nonradiative SNR evolution by \citet{true99}, who predict analytically
and numerically the expected forward blastwave and reverse shock
positions throughout a remnant's lifetime, for a corrobrative
explanation.  Based on the fact that \gll is likely 1616~yr old, and
the expansion parameters we find corresponding to that age, we
estimate roughly that the SNR reverse shock radius is currently
between 0.5 and 0.8 times the forward blastwave shock radius, assuming
a typical ejecta mass for Type~II SN of 3$-$5 $M_{\odot}$.  The ratio
of PWN to SNR diameter is $\sim0.28$, so the reverse shock would not be
expected to have reached the PWN yet.  This agrees with what we
observe in Figure~\ref{fig:dif}, as well as conclusions outlined in
\citet{tamr02} based on the hydrodynamical simulations of
\citet{vand01}.  Furthermore, we refer to \citet{chev82}, who predicts
that the initial expansion phase of a SNR with a red supergiant
progenitor can be described by a self-similar solution with a value of
$m=0.9$, as long as both the circumstellar material and the stellar
envelope density distributions are power-laws in radius.  This is
considerably higher than our $m$ estimates, based on $t=1616$ yr, of
roughly 0.48 to 0.68, which suggests that the transition from the
initial phase to the Sedov phase is well under way. 

\subsection{Distance Estimate}
The distance to \gll is estimated by considering its expansion and
angular size.  Given the relations $v=m R/t$ and $R=\theta d$, where
$v$ is the shell's spatial velocity, $\theta$ is the angular radius,
and $d$ is the distance to the remnant, it can be seen that 
\[d=v\left( \frac{\Delta t}{\theta \cdot \Delta R/R}\right).\]
To find the velocity of the shell, we consider the Mach number of the
SNR shock front $M=v/c_s$, where $c_s=\sqrt{\gamma p/\rho}$ is the
sound velocity \citep{long94}.  Here, $\gamma=5/3$, and the ratio of
pressure to particle mass density is given by $p/\rho=kT_1/\mu m_p$,
where $T_1$ is the temperature of the surrounding material and $\mu
m_p$ is the mean mass per particle ($\mu=0.6$ for cosmic abundances,
$m_p$ is 
proton mass) \citep{reyn94}.  \citet{long97} quotes $T_2/T_1=5M^2/16$
for a strong shock in an ideal gas.  We measure the temperature behind
the shock $T_X \simeq 7 \times 10^6$ K from X-ray spectral fits
\citep{robe03}; however, $T_2$ is the ion temperature and if the
electrons are not in full thermal 
equilibrium with the ions, the observed spectrum may underestimate the
shock temperature, and our distance estimate will be too small by a
factor of $\sqrt{T_2/T_X}$ \citep{bork01}.  Combining the above
information we find 
\[\frac{T_2}{T_1} = \frac{5}{16} \left(\frac{v^2}{5kT_1/3\mu
m_p}\right)\] 
which gives a lower bound on the distance estimate:
\[d=\left( \frac{16kT_2}{3\mu m_p}\right)^{1/2} \left( \frac{\Delta t}
{\theta \cdot \Delta R/R}\right) \gtrsim 3
\sqrt{\frac{T_2}{T_X}} \left(
\frac{0.0071}{\Delta R/R} \right)  \mathrm{kpc}.\]
\citet{gree88} previously estimated a minimum distance of $\sim5$~kpc
to the remnant based on its \ion{H}{1} spectrum.  Fits to the X-ray
spectrum with the NPSHOCK model of \citet{bork01} suggest that the
electrons are near equilibrium; therefore, the distance derived
assuming total equilibrium should be very close to that at near
equilibrium, and, hence, not much greater than the minimum \ion{H}{1}
distance.  

\section{CONCLUSIONS}

Based on radio interferometric images of SNR \gll we have made a
simple measurement of the outer shell expansion and found a mean rate
of $0\farcs057 \pm 0\farcs012$/yr from 20-cm data, and $0\farcs040 \pm
0\farcs013$/yr from 6-cm data.  If we compare the expected age of
G11.2$-$0.3, determined by our measurements, with the characteristic
age of its associated pulsar PSR~J1811$-$1925, we find an order of
magnitude discrepancy; our result further strengthens the growing body
of evidence linking \gll with the historical SN of 386~AD.  The
evolutionary status of this SNR appears to be pre-Sedov, a conclusion
that agrees with other observational evidence, as well as theoretical
arguments.  We also estimate the distance to the remnant based on its
X-ray shock velocity to be $\gtrsim3$ kpc and find it consistent with
previously published results.

\acknowledgments
We wish to thank S. M. Ransom, F. P. Gavriil, V. M. Kaspi, M. Lyutikov
and S. P. Reynolds for their helpful comments and suggestions.  The
National Radio Astronomy Observatory is a facility of the National
Science Foundation operated under cooperative agreement by Associated
Universities, Inc.

\begin{deluxetable}{lcccc}
\tablewidth{350pt}
\tablecaption{VLA observing parameters for epoch 2 \label{tab:obs} } 
\tablehead{
\colhead{Observing} & \colhead{Array} & \colhead{Frequencies} &
\colhead{Bandwidth} & \colhead{Time on} \\ 
\colhead{Date} & \colhead{Config.} & \colhead{(MHz)} &
\colhead{(MHz)} & \colhead{Source (min)} }
\startdata
2001 Jun 26 & CnB & 1465 & 25 & 63 \\
2001 Jun 26 & CnB & 4835, 4885 & 25 & 81 \\
2001 Aug 03 & C & 1465 & 25 & 60 \\
2001 Aug 03 & C & 4835, 4885 & 25 & 66 \\
2001 Sep 24 & DnC & 1465 & 25 & 92 \\
2001 Sep 24 & DnC & 4835, 4885 & 25 & 99 \\
2002 May 24 & BnA & 1465 & 25 & 98 \\
2002 May 24 & BnA & 4835, 4885 & 25 & 118 \\
\enddata
\end{deluxetable}

\end{document}